\documentclass[twocolumn,prl,showpacs]{revtex4}

\usepackage{graphicx,epsfig,amssymb}

\begin{document}

\title{Spectral fluctuations effects on conductance peak height 
statistics in quantum dots}
 
\author{K. Held, E. Eisenberg, and B. L. Altshuler}
\affiliation{Physics Department, Princeton University, Princeton, NJ 08544\\
 NEC Research Institute, 4 Independence Way, Princeton, NJ 08540}

\begin{abstract}
Within random matrix theory for  quantum dots, both the dot's one-particle
eigenlevels and the dot-lead couplings are statistically
distributed. While the effect of the latter on the conductance is obvious
and has been taken into account in the literature, the statistical
distribution of the one-particle eigenlevels is generally replaced by a 
picket-fence spectrum. Here we take the random matrix theory eigenlevel
distribution explicitly into account and observe significant deviations
in the conductance distribution  and magnetoconductance of closed quantum dots
at experimentally relevant temperatures.
 

\end{abstract}

\date{\today}

\maketitle

The universal statistical fluctuations observed at the low-energy part of the
spectrum of quantum systems whose associated classical dynamics are
chaotic can be described by random matrix theory (RMT). 
This type of description  can be 
justified for diffusive quantum dots
and quantum dots with irregular shapes
\cite{Alhassid01} which makes quantum dots a particular example 
for the study of RMT fluctuations.
While in open quantum dots (which have a strong
dot-lead coupling) the
electron-electron interaction effects are mostly 
neglected,  in closed quantum dots this interaction leads to the Coulomb 
blockade (see \cite{CBreview} for a review):
The low-temperature conductance is heavily suppressed due to the large
charging energy for adding an electron to the quantum dot,
 except for
the Coulomb blockade peaks at which the potential of the quantum dot is 
adjusted such that  $N$ and $N+1$ electrons in the dot
have the same energy. The RMT approach \cite{Jalabert92,Alhassid01} successfully described
the mesoscopic fluctuations of these
Coulomb blockade peaks, i.e., the statistical distribution of their
height $P(G^{\rm max})$  and its dependence upon magnetic field 
\cite{ChangFolk96}. On the other hand, 
recent improved experiments show significant deviations from the RMT 
prediction, suggesting that interaction effects beyond charging should be 
considered as well. In particular, dephasing of the single-particle 
states due to interactions modifies the conductance peak height statistics
(see \cite{Alhassid01,Aleiner01} and references therein).
In a recent experiment, Patel {\it et al.} \cite{Patel98}
found that the statistical distribution has a smaller ratio of
standard deviation  to mean peak height 
$\sigma(G^{\rm max})/\langle G^{\rm max}\rangle$
than predicted by RMT \cite{Alhassid98} and attributed this
to dephasing effects. In another experiment, Folk {\it et al.} \cite{Folk00} 
measured the change of the conductance in a magnetic field $B$
\begin{equation}
\alpha =\frac{\langle G^{\rm max} \rangle_{B\ne 0}-\langle G^{\rm max} 
\rangle_{B=0}}{\langle G^{\rm max} \rangle_{B\ne 0}},
\label{Eq:alpha}
\end{equation}
as a probe of dephasing times. This is the closed dot analog of
the weak localization magnetoconductance
which had proven to be an effective measure for open dots
dephasing times \cite{Huibers98}.
It was pointed out that  $\alpha=1/4$ 
as long as the transport is dominated by 
elastic scattering \cite{Alhassid98,Beenakker00}. 
Therefore, any deviation of the measured
$\alpha$ from $1/4$ was considered an indication for dephasing.

In this paper we discuss the effects of spectral fluctuations 
of the RMT one-particle
eigenlevels on the statistical distribution $P(G^{\rm max})$
and the weak localization correction 
$\alpha$. 
Previous works 
\cite{Alhassid01,Jalabert92,Alhassid98,Beenakker00,Eisenberg01}
have generally considered a 
picket fence spectrum, i.e., a rigid level spacing
between successive eigenlevels in the quantum dot,
for the calculation of the conductance. This ignores the effect of 
spectral eigenlevel fluctuations. The picket fence spectrum
is a good approximation for both very high temperatures
and very low temperatures \cite{Alhassid01}, 
and a  comparison of $P(G^{\rm max})$
with full RMT statistics and a picket fence spectrum without spin-degeneracy
at three temperatures showed only minor deviations \cite{Alhassid98b}.

In the present paper, we study the full RMT statistics in detail
with and without spin-degeneracy,  and
find significant differences compared to the  picket fence spectrum,
in particular in an
experimentally relevant regime $k_B T\lesssim \Delta$.
The spectral fluctuations lead to lower values of   $\alpha$ 
than  $1/4$ such that this value is not universal,
even in the absence of any dephasing mechanism. 
One therefore has to be careful while using $\alpha$ as a probe for dephasing 
in this temperature regime.

Within the constant interaction model, 
the conductance of a quantum dot is given by 
the  formula 
\cite{Beenakker91}
\begin{equation}
G=\frac{e^2}{kT}\sum_{p=1}^\infty
\frac{\Gamma^L_i\Gamma^R_i}{\Gamma^L_i\!+\!\Gamma^R_i}
P_{\rm eq}(N)P(E_i|N)[1\!-\!f(E_i\!-\!\mu)]
\label{Eq:G}
\end{equation}
where $\Gamma^{L(R)}_i$ is the tunneling rate between the $i$th 
one-particle eigenlevel 
of the dot and the left (right) lead, $P_{\rm eq}(N)$ is the 
equilibrium probability to find $N$ electrons in the dot
with the Coulomb blockade allowing for $N$ and $N+1$ electrons, 
$P(E_i|N)$ is the canonical probability to have the 
level $i$ occupied given the 
presence of $N$ electrons in the dot, and $f(E)$ is the Fermi function
at the effective chemical potential $\mu$, which includes the charging energy.
In a typical experimental situation, the charging energy is much large than
temperature, and thus only one term contributes to the sum over $N$.
In Eq.~(\ref{Eq:G}), $\Gamma^{L(R)}_i$ is Porter-Thomas distributed
in the Gaussian orthogonal ensemble
(GOE)  and  Gaussian unitary ensemble (GUE) 
without and with a magnetic field, respectively, and
the eigenlevel energies $E_i$ obey the respective RMT distribution
\cite{Alhassid01}. In contrast, the picket fence spectrum has
$E_{2i}=E_{2i-1}=i \Delta$ in the case of spin-degeneracy and
$E_{i}=i \Delta/2$ without spin-degeneracy.
The first term in the sum $\Gamma^L_i\Gamma^R_i/(\Gamma^L_i+\Gamma^R_i)$
depends only on the eigenfunctions of the dot, and thus is uncorrelated 
with the spectrum within the RMT approach. The ensemble average of this term in the 
absence (GOE) or presence (GUE) of a magnetic field
is  
\begin{equation}
\left\langle\!\!\left\langle\frac{\Gamma^L_i\Gamma^R_i}{\Gamma^L_i+\Gamma^R_i}\right\rangle\!\!\right\rangle=
\left\{ \begin{array}{l l} 
1/4 & \;\; \mbox{GOE}\\  
1/3 & \;\; \mbox{GUE}
\end{array}
\right. .
\end{equation}
 This yields the value $\alpha=1/4$ if the weights
 $P(E_i|N)$ are the same for both ensembles. This should be the case
in the low temperature regime $k_B T\ll \Delta$ since only
one level $E_0$ contributes with maximal weight, $P(E_i|N)\approx\delta_{i0}$. 
In general, the main
contribution to the sum comes from $O(k_BT/\Delta)$ levels around the 
Fermi energy which gives the same contribution at large temperatures
$k_B T\gg \Delta$ for the GOE and GUE,
$\alpha=1/4$ in this regime as well. 

\begin{figure}[tb]
\includegraphics[width=3.25in]{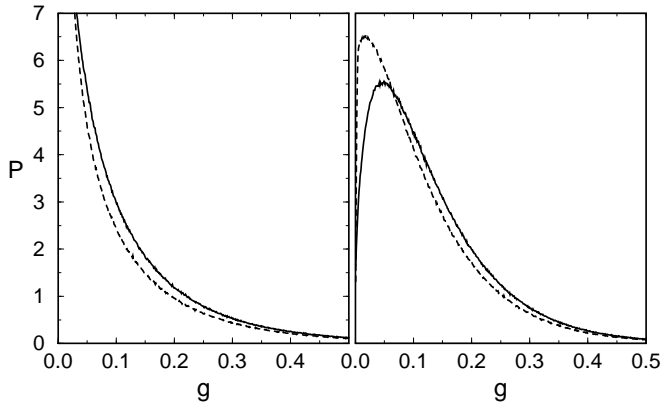}
\vspace{-.5cm}

\caption{Probability distribution $P(g)$  of the dimensionless
closed dot conductance $g$ defined by $G^{\rm max}\!= \!\frac{e^2}{\hbar}\frac{\hbar\bar\Gamma}{k_B T}  g$
at $k_B T=0.2 \Delta$ in the presence of spin-degeneracy
(left:  GOE; right: GUE; solid line:  RMT spectral fluctuations; dashed line:
picket fence)
} 
\label{alpha}
\end{figure}

However, for $k_B T \lesssim \Delta$,
the probability  to have more than one level in an energy window 
$k_BT$ around the 
Fermi energy is increased for the RMT eigenlevel distribution compared to
the picket-fence spectrum.
These additional levels enhance the conductance.
Since there are more close-by levels for the GOE case, due to the weaker
level repulsion, the GOE conductance is enhanced more, 
and $\alpha$ is suppressed.

A second important  effect is the optimization
of the chemical potential for the Coulomb blockade peak.
This effect was generally ignored, as it is technically
cumbersome to consider, and is not significant for both very low and very 
high temperatures. Disregarding this effect means that a theorist 
optimized the chemical potential w.r.t. the averaged conductance, 
instead of optimizing for every realization as in the experiment. 
Whenever there is a close-by level, the position of the
peak is shifted to optimize the contribution from both levels. 
Typically, a level with very low tunneling rates (and, thus,
suppressed conductance peak) would get enhanced significantly
by contributions from
its neighbors. If the tunneling rate of 
a neighboring level is much higher,
the peak position $\mu^{\rm max}$ is shifted towards it. 
As the distribution of level spacings is different depending
on the existence of magnetic field, this enhancement mechanism is again more 
effective in the absence of magnetic field (GOE), where probabilities
of small spacing and of small conductances are higher. 
Thus, this effect  which was neglected in \cite{Alhassid98b} 
suppresses $\alpha$ even further.

\begin{figure}[tb]
\includegraphics[width=3.2in]{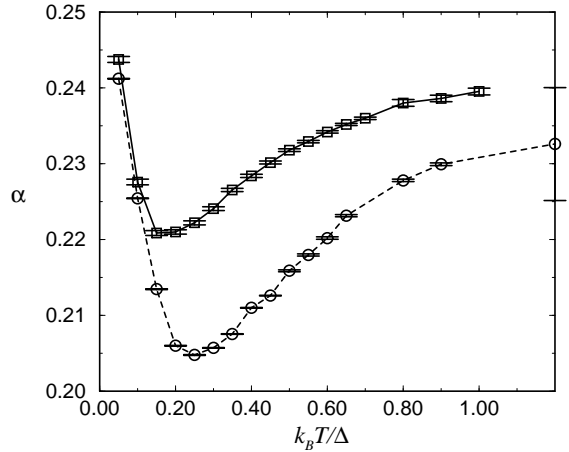}
\vspace{-.5cm}

\caption{Magnetoconductance $\alpha$ vs. $k_B T/\Delta$ for the spin-degenerate case
(dashed line) and without spin-degeneracy (solid line). Taking into
account the RMT spectral fluctuations, $\alpha$ is reduced from its
``universal'' value $\alpha=1/4$, in particular in the experimental relevant
regime $0.1 \Delta < k_B T < 0.8 \Delta$.} 

\end{figure}

We evaluated the sum (\ref{Eq:G}) numerically by drawing $\Gamma^{L(R)}_i$
from the Porter-Thomas distribution and $E_i$ according to the
Wigner-Dyson distribution.
Levels within a window of $\pm 4 k_B T$ around the Fermi energy have been
taken into account and the Fermi energy $\mu$ in Eq.~(\ref{Eq:G})
has been adjusted to yield $G^{\rm max}$ for every realization. 

Figure 1 compares the probability distribution $P(G^{\rm max})$ 
for a picket fence spectrum vs. the full RMT level statistics. 
As explained above, RMT spectral fluctuations  enhance the conductance. 
In particular, the probability to have a very low $G^{\rm max}$ is 
reduced and the probability to have 
an intermediate  $G^{\rm max}$ is enhanced. 
The reason for the reduction is that a very low  $G^{\rm max}$
requires $\Gamma^L$ or $\Gamma^R$ in Eq.~(\ref{Eq:G}) to be low. 
RMT spectral fluctuations
enhance the contributions from close-by levels, 
which typically do not have a low
value of $\Gamma^{L(R)}$ at the same time. Thus, the peak position of $\mu$ 
is shifted towards a close-by level 
and the conductance occurs through both levels.
Notably, the effect of phase-breaking inelastic scattering processes
leads to similar changes \cite{Eisenberg01}.

Deviations of $\alpha$ from the ``universal'' value $1/4$ have been
interpreted as being a result of dephasing. While dephasing would certainly
suppress $\alpha$, we note here that in the regime 
$k_B T\lesssim \Delta$, the spectral fluctuation effects discussed above, lead to 
a similar effect.
In Figure 2 we present the results for $\alpha$ as a function of the scaled
temperature $k_BT/\Delta$, for both spin-degenerate spectrum and the case of
broken symmetry. While the effect seems to be small, 
one should keep in mind that
in the low temperature regime, even very strong dephasing does not suppress 
$\alpha$ to zero \cite{Beenakker00}, and thus the correction due
to spectral fluctuations is comparable with or even larger than 
the effect of dephasing \cite{Beenakker00,Eisenberg01}. 
One should therefore cautiously use $\alpha$ as a probe
of dephasing in this regime.

In conclusion, we have shown that RMT spectral fluctuation effect the
probability distribution function $P(G^{\rm max})$, leading to 
non-negligible deviations in measurable quantities in the regime
$0.1 \Delta <k_B T<0.8\Delta$. In particular,  the
weak localization correction $\alpha$ which was recently used as 
a probe of dephasing in closed quantum dots is affected:
$\alpha$ is different from $1/4$ and, moreover, turns out to
be temperature dependent, even in the absence of 
dephasing. At low temperatures, $\alpha$ is reduced down to 
$\alpha \approx0.2$, which can be  below the lower limit of a 
picket fence model {\em with} dephasing. 
This should also be taken into account while analyzing the ongoing experiments
aimed at measuring dephasing times in closed dots in the low 
temperature regime $k_B T\lesssim \Delta$. 

Finally, we would like to note that during the completion of the present
paper some of the results have been independently arrived at
in \cite{Alhassid02}.


\end{document}